\documentclass[final,5p,times,twocolumn]{elsarticle}

\usepackage{amssymb}
\usepackage{amsmath}

\usepackage{url}
\usepackage{amsmath,graphicx}
\usepackage[dvipsnames]{xcolor}
\usepackage{amsfonts}
\usepackage{bm}
\usepackage{hyperref}
\usepackage{cleveref}
\usepackage{multirow}
\usepackage{lipsum}
\usepackage{booktabs}
\usepackage{bm}
\usepackage{multirow}
\usepackage{makecell}
\usepackage{enumitem,kantlipsum}

\makeatletter
\def\ps@pprintTitle{%
  \let\@oddhead\@empty
  \let\@evenhead\@empty
  \let\@oddfoot\@empty
  \let\@evenfoot\@empty}
\makeatother

\journal{Bioacoustics}

\begin{document}

\begin{frontmatter}



\title{On feature representations for marmoset vocal communication analysis}

\author[label1,label2]{Eklavya Sarkar\texorpdfstring{\corref{cor1}}}
\ead{eklavya.sarkar@idiap.ch}
\cortext[cor1]{Corresponding author.}
\affiliation[label1]{organization={Idiap Research Institute}, city={Martigny}, postcode={1920}, country={Switzerland}.}
\affiliation[label2]{organization={Ecole polytechnique fédérale de Lausanne}, postcode={1015}, country={Switzerland}.}

\author[label3]{Kaja Wierucka}
\ead{kwierucka@dpz.eu}
\affiliation[label3]{organisation={Behavioral Ecology and Sociobiology Unit, German Primate Center, Leibniz Institute for Primate Research}, city={Göttingen}, postcode={37077}, country={Germany}.}

\author[label4]{Alexandra B. Bossard}
\ead{alexandra.bosshard@uzh.ch}
\affiliation[label4]{organization={Department of Comparative Language Science, Institute of Evolutionary Anthropology, and Center for the Interdisciplinary Study of Language Evolution (ISLE), University of Zurich}, postcode={8050}, country={Switzerland}.}

\author[label4]{Judith Burkart}
\ead{judith.burkart@aim.uzh.ch}

\author[label1]{Mathew Magimai.-Doss}
\ead{mathew@idiap.ch}

\begin{abstract}
The acoustic analysis of marmoset (\textit{Callithrix jacchus}) vocalizations is often used to understand the evolutionary origins of human language. Currently, the analysis is largely carried out in a manual or semi-manual manner. Thus, there is a need to develop automatic call analysis methods. In that direction, research has been limited to the development of analysis methods with small amounts of data or for specific scenarios. Furthermore, there is lack of prior knowledge about what type of information is relevant for different call analysis tasks. To address these issues, as a first step, this paper explores different feature representation methods, namely, HCTSA-based hand-crafted features Catch22, pre-trained self supervised learning (SSL) based features extracted from neural networks trained on human speech and end-to-end acoustic modeling for call-type classification, caller identification and caller sex identification. Through an investigation on three different marmoset call datasets, we demonstrate that SSL-based feature representations and end-to-end acoustic modeling tend to lead to better systems than Catch22 features for call-type and caller classification. Furthermore, we also highlight the impact of signal bandwidth on the obtained task performances.
\end{abstract}


\end{frontmatter}



\section{Introduction}
\label{sec:intro}
The advancements in human speech processing have accelerated and impacted research in non-human communication, such as bioacoustics, i.e. the study of animal sounds. Common marmosets (\textit{Callithrix jacchus}) have recently gained prominence as a valuable research model among non-human primates. This is primarily due to their exceptional vocal abilities, which are rooted in their highly complex social behavior and cooperative breeding system \cite{Eliades2017, burkart2022}. They possess extensive vocal repertoires used in various social situations \cite{Agamaite2015, bezerra2008}, and their vocalizations have the capacity to encode a wide range of information, such as population, group affiliation, sex, and even individual identity \cite{Zurcher2017, Jones1993, Newman1992, Rukstalis2005, Norcross1993, Phaniraj2023}. These vocalizations are not limited to simple tonal signals but also encompass complex calls with multiple frequency components, some of which are within the ultrasonic range \cite{Bakker2018}. Moreover, marmosets have been observed to exhibit remarkable vocal adaptability. They can alter the duration \cite{brumm2004}, intensity \cite{brumm2004, Eliades2012, Pomberger2020}, complexity \cite{Pomberger2018}, or timing \cite{Roy2011, Pomberger2018} of their calls, even when faced with disruptions in their environment that occur after the initiation of a call \cite{Pomberger2020}. While these properties make marmosets an intriguing subject for the study of communication processes, they also pose a significant challenge when attempting to automate the analysis of their vocalizations. The literature on automatic marmoset vocalization analysis is relatively sparse.

Turreson et al. compared different classification methods for marmoset `call-type' classification using linear prediction coefficients as feature representation, and found that on a small data setup of 30 samples per call-type, k-NN, SVM and optimal path forest algorithms yield better performance than multilayer perceptron, Adaboost, and logistic regression \cite{lpc_ML_marmoset_2016}. Wisler et al. investigated different feature representations, namely, audio features (statistics based on energy entropy, signal energy, zero crossing rate, spectral rolloff, spectral centroid, and spectral flux), mel-frequency cepstral coefficients (MFCCs), and Teager energy operator-based features for marmoset vocalization and call-type detection \cite{wisler16_interspeech}. On a synthetic dataset, created by taking a small set of calls and augmenting it with background noise and acoustic events, it was found that feature level combination leads to better performance. Verma et al. investigated discovering of different patterns in marmoset calls through unsupervised learning. Specifically, they developed an HMM-based approach to segment and cluster marmoset vocalizations into discrete units through multi-resolution and multi-rate analysis of the signal \cite{verma17_interspeech}. In \cite{cas_data}, it was demonstrated that marmoset vocalizations and call-types can be better detected/classified by feeding statistics of log-mel-filter bank energies as input to recurrent neural networks, when compared to feeding it to SVM or multilayer perceptrons. In the scenario of analyzing recordings obtained from a pair of marmosets, \cite{Oikarinen_joint_vad_caller_call_marmoset_jasa_2019} investigated a deep learning approach where a spectrogram was fed as input to a convolutional neural network to jointly perform vocalization, detection, call type classification and caller detection. It was found that joint modeling yielded better performance than training systems individually for each task in this scenario. Recently, Highly Comparable Time-Series Analysis  (HCTSA) features have been used to model source (caller) identification through an Adaboost-based hierarchical approach for marmosets \cite{Phaniraj2023}, as well as for 14 mammalian species \cite{Wierucka2024.04.14.589403}. In the bioacoustics field, breakthroughs in self-supervised learning (SSL), which leverages unlabeled data by creating surrogate labels from the data's inherent structure, has led to works which explore birdsong detection \cite{cola_paper} and bioacoustic event detection \cite{bioacoustic_event} by pre-training with a contrastive learning approach. In that direction, a study using different SSLs pre-trained on human speech, demonstrated that neural embeddings extracted in such a framework can also distinguish marmoset callers \cite{Sarkar_INTERSPEECH_2023}.

However, in the existing works, there are three main limitations. First, most of the studies have been carried out on small data sets. Second, these studies have been conducted on datasets intended for specific scenarios. Due to a lack of validation, it is unclear whether the methods studied on one dataset would scale to another. Third, there is limited prior knowledge about what type of information is relevant for different call analysis tasks. There is a need to overcome these limitations to advance the development of automatic analyses of marmoset vocalizations. The present paper is a step in that direction with a specific focus on feature representations for automatic marmoset call analyses, where we investigate three prominent feature representation methods, namely, (a) hand-crafted features, (b) self-supervised learning-based representations, and (c) end-to-end acoustic modeling, on three different marmoset call datasets and three different tasks (call type, caller identity, and caller sex classification).

The paper is organized as follows. \Cref{sec:methodology} presents the different datasets, tasks, and investigated feature representations. \Cref{sec:exp_study} and \ref{sec:analysis} present the studies and analysis of the results respectively. Finally \cref{sec:conclusion} concludes the paper. 

\section{Methodology}
\label{sec:methodology}

\subsection{Datasets and tasks}
\label{sec:datasets_protocol}

We conduct investigations on three different marmoset datasets, denoted as $D_1$, $D_2$, and $D_3$, respectively. $D_2$ and $D_3$ contain vocalizations produced by adult individuals, while $D_1$, InfantMarmosetsVox, originates from infant marmosets~\cite{Sarkar_INTERSPEECH_2023}. Consequently, $D_1$ is expected to encompass different call types, likely characterized by higher frequencies compared to those in $D_2$ and $D_3$. Furthermore, $D_2$ and $D_3$ are gathered from the same colony, while $D_1$ was obtained from a different one. All the datasets consist of audio recordings of marmosets vocalizations segments, collected and hand-labeled with the start and end time by experienced researchers. In addition to call-type and caller identity annotations of each vocalization provided for all three datasets, $D_1$ and $D_2$ also include information about the sex of the vocalizing individual. For more details regarding the datasets, the reader is referred to \ref{sec:data_description}.

\begin{figure}[!htb]
  \centering
  \includegraphics[width=\linewidth]{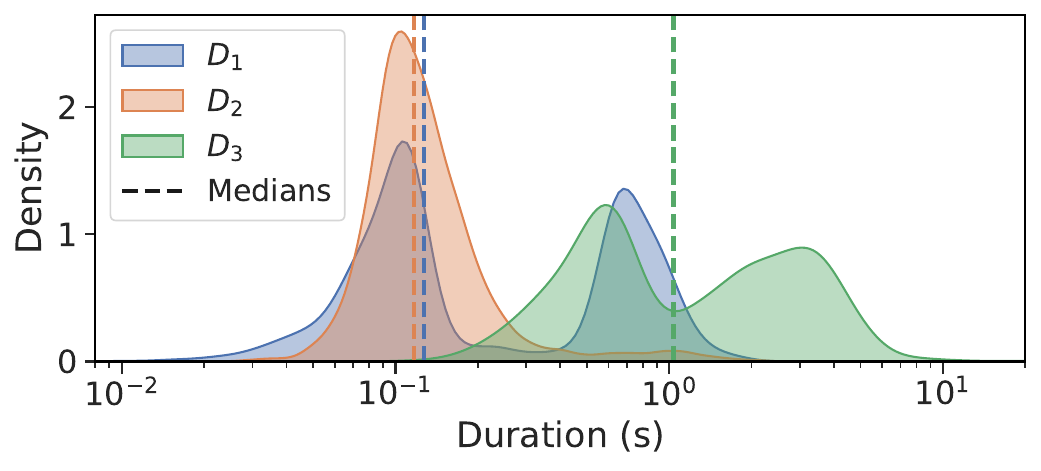}
  \caption{Log distribution of vocalization lengths per dataset. The medians are calculated over the entirety of each dataset.}
  \label{fig:durations}
\end{figure}
We discard any segments labeled as `silence' and `noise', and only keep the vocalization segments. The log distribution of the vocalization lengths of the three datasets is presented in \Cref{fig:durations}. We can observe that $D_1$ has the shortest median vocalization length at 127 ms, with $D_2$ and $D_3$ at 175 and 1037 ms respectively. Based on the given annotations, we define multi-class tasks, specifically call-type, caller, and sex classification, henceforth referred to as CTID, CLID, and SID respectively. \Cref{table:datasets} gives the number of vocalization segments $S$, their total duration length $L$, the native sampling rates, as well as the number of classes $n_c$ for each task across datasets.

\begin{table}[!htb]
\centering
\caption{$S$ indicates the number of data samples, $L$ the sum of all vocalizations segment durations (in minutes), and SR the native sampling rate of the given data (kHz). $n_\text{task}$ is the number of classes of each task-dataset permutation.}
\begin{tabular}{lcccccc}
\toprule
$\bm{\mathcal{D}}$ & $\bm{S}$ & $\bm{L}$ & \textbf{SR} & \bm{$n_\text{CTID}$} & \bm{$n_\text{CLID}$} & \bm{$n_\text{SID}$} \\
\midrule
$D_1$ \cite{Sarkar_INTERSPEECH_2023} & $73$K & $464$ & 44.1 & 11 & 10 & - \\
$D_2$ \cite{Alex_MA_thesis} & $14$K & $37$ & 300 & 7 & 8 & 2 \\
$D_3$ & $5$K & $138$ & 125 & 12 & 8 & 2 \\
\bottomrule
\end{tabular}
\label{table:datasets}
\end{table}

\subsection{Feature representations}
\label{sec:feat_representation}
We investigate the following feature representations:

1) \textit{Hand-crafted features}: Highly Comparable Time-Series Analysis (HCTSA) is an interpretable signal processing-based framework that has been demonstrated to be useful for diverse time series application domains~\cite{hctsa_2013}. In this framework, a set of 7700 features are extracted by characterizing the signal by different time series analysis methods, such as, linear correlation, modeling fitting (e.g., autoregressive moving average analysis, GARCH), wavelet analysis, extraction of information theoretic measures, which then is combined with feature selection to build statistical models for the end task. In the literature, these features have been investigated for behavioural birdsong discrimination~\cite{birdsong_hctsa}, automated acoustic monitoring of ecosystems~\cite{sethi2020automated}, as well as marmoset caller identification~\cite{Phaniraj2023}. One of the challenges of HCTSA approach is computational complexity and involves an evaluation of many similar features. In a recent work, CAnonical Time-series CHaracteristics (Catch22) features, a subset of the HCTSA feature set has been proposed which exhibit a strong performance across 93 real-world time-series classification problems, but are also minimally redundant~\cite{lubba2019catch22}. In this work, we investigate the Catch22 features, denoted as C22.

2) \textit{Pre-trained self-supervised learning (SSL) based features}: Inspired from the recent study presented in~\cite{Sarkar_INTERSPEECH_2023}, we investigate the use of feature representations extracted from pre-trained SSL neural networks trained on human speech for marmoset call analysis. We extend the investigations from caller detection to call type, caller ID and sex classification. Furthermore, contrary to the previous work~\cite{Sarkar_INTERSPEECH_2023}, which focused only on the last transformer layer representation, in this work we investigate representations obtained from all the transformer layers to gain insight which level of layer representations are informative for marmoset call analysis.

3) \textit{End-to-end acoustic modeling}: With advances in deep learning, acoustic modeling approaches have emerged in speech and audio processing where raw signal can be modeled to learn task-dependent information from the signal in an end-to-manner with minimum prior knowledge~\cite{palaz_phoneme_2013, Trigeorgis16, Zazo16, hannah_verif_2018}. Such approaches hold potential for advancing marmoset call analysis, as they could help not only in addressing the lack of reliable task-dependent prior knowledge challenge, but also in gaining insight into the task relevant acoustic information learned by such trained networks through analysis~\cite{hannah_verif_2018, hannah_understanding_2019, Palaz_SPEECHCOMMUNICATION_2019}. The insight gained could then be further validated through linguistic studies. Motivated by these aspects, we investigate this approach.

A sub-challenge that arises when analyzing marmoset calls is the range of frequency information to be modeled. More precisely, the fundamental frequencies (typically corresponding to the peak frequency) of adult marmoset vocalisations span a range of 6-13 kHz, depending on the \emph{call-type}~\cite{Agamaite2015}. However, as can be seen in \Cref{table:datasets}, datasets are collected at varying sampling frequencies. Furthermore, the SSL neural networks are typically pre-trained on speech signal of 8 kHz bandwidth (i.e., 16 kHz sampling frequency). As part of the investigation, we thus also study the impact of sampling rate (SR) on marmoset call analysis tasks.

\section{Experimental Study}
\label{sec:exp_study}

\subsection{Systems}
For each task, we divided all datasets into training, validation, and test sets, named \textit{Train}, \textit{Val}, and \textit{Test} respectively, following a 70:20:10 split ratio, in order to train models on a sufficiently large number of samples, while ensuring sufficient data points for model evaluation and validation. \textit{Train} is used to train the models, \textit{Val} to tune any hyperparameters, and \textit{Test} to evaluate the trained models on unseen data. We then developed the following systems for each task on each dataset to investigate the aforementioned feature representations:

1) We used \textit{pycatch22} to extract a feature vector $\bm{x} \in \mathbb{R}^{1 \times D}$ (denoted as C22) for each utterance, where $D=24$, and feed it to a multilayer perceptron (MLP) with three hidden layers of 128, 64, and 32 number of hidden units, respectively. The classifier is trained for $30$ epochs, using a batch size $16$ and learning rate $\eta=1e-3$.

2) As it is challenging to investigate all the different types of pre-trained SSL feature representations across all tasks and datasets, we simply chose WavLM~\cite{wavlm_paper}, as it was found to yield strong performance on the task of marmoset caller detection \cite{Sarkar_INTERSPEECH_2023}, been found to scale well to different human speech processing tasks in the SUPERB challenge~\cite{yang21c_interspeech}. For each layer, we extracted frame-by-frame variable-length feature representations $\bm{x} \in \mathbb{R}^{N \times D}$, where $D=768$ and $N$ the variable number of frames (contingent on the vocalization length). We then converted these embeddings into utterance-level fixed-length representations $\bm{f}_{\mu\sigma} \in \mathbb{R}^{1 \times 2D}$ (denoted as WLM), by computing and concatenating the first and second order statistics across the frame axis on the extracted features. An MLP of same three layer architecture as \textit{C22} is then trained with the fixed length feature as input.

3) We trained a convolutional neural network (CNN) based end-to-end acoustic modeling system (denoted as E2E) that takes a raw waveform as input and classifies to the output classes. Following the literature in speech processing~\cite{Dubagunta_ICASSP-2_2019, Nallanthighal_NEURALNETWORKS_2021, Purohit_ICASSP_2023}, the E2E system consists  of four convolution layers followed by an adaptive pooling layer and two hidden layers. The E2E system is optimized with a cross-entropy cost function with an early stopping criteria. Further details of the architecture are provided in the ~\ref{sec:cnn_arch}.

In the case of \textit{C22}, we developed systems at native sampling frequency and downsampled acoustic signals: 16 kHz for $D_1$, 60 and 16 kHz for $D_2$, and 60 and 16 kHz for $D_3$. In the case of WLM, we developed systems with signals downsampled to required pre-training sampling rate of 16 kHz. For E2E system, $D_2$ and $D_3$ signals were downsampled to 60 and 16 kHz. 
To evaluate the systems we used Unweighted Average Recall (UAR) as the metric to account for any class imbalance.

\subsection{Results}
\begin{figure}[!htb]
  \centering
  \includegraphics[width=\linewidth]{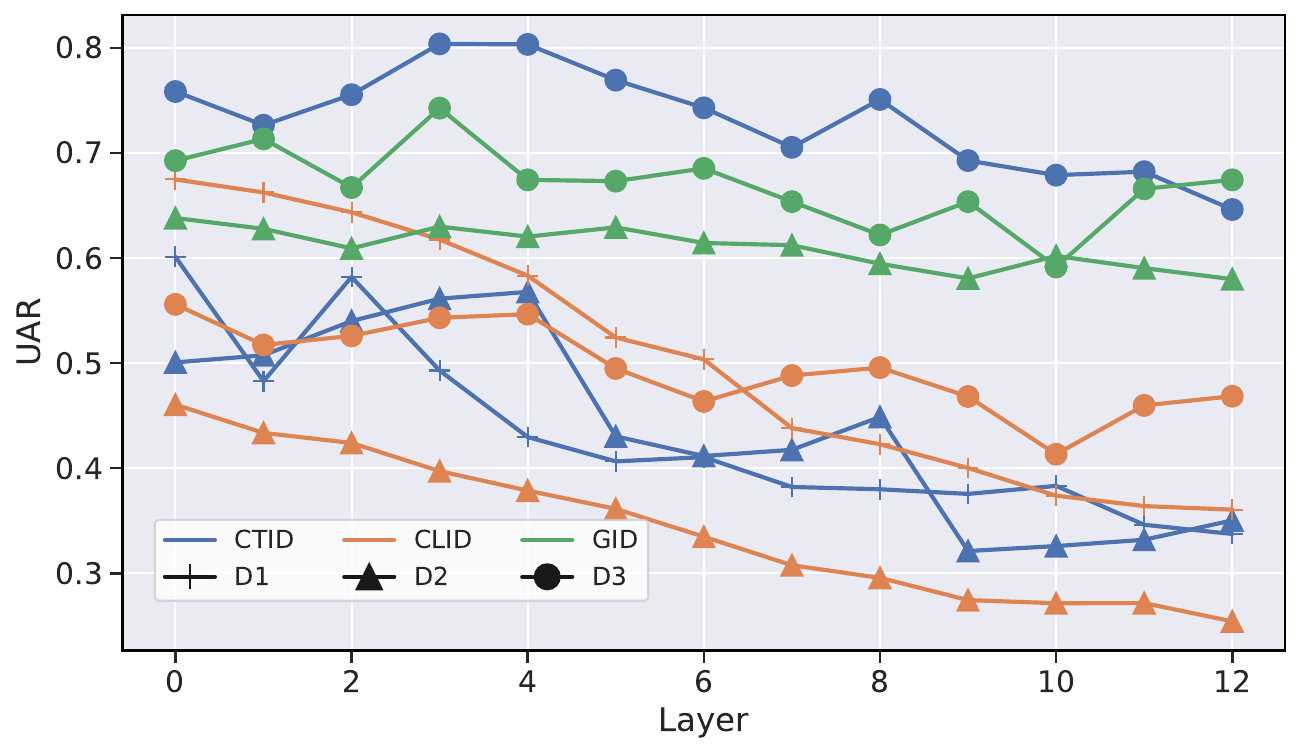}
  \caption{Layer-wise UAR scores for WLM for all tasks and datasets. The layers follow the same indexing as \cite{wavlm_paper}.}
  \label{fig:combined_layers}
\end{figure}

\Cref{table:uar_results} shows the performances of systems based on different feature representations. For the sake of clarity, only the best layer and worst layer performances are reported for WLM. \Cref{fig:combined_layers} presents the layer-wise performances for all tasks on all datasets for WLM. Note that layer 0 corresponds to the output embedding of the CNN encoder, where as the other 12 refer to the outputs of the transformer encoder layers. The performances are all above chance level, i.e. 100/$n_c$, for all systems.

\begin{table}[!htb]
\centering
\caption{UAR scores on \textit{Test} on features $\mathcal{F}$. WavLM's best and worst layer's score is given. For each dataset, the best score across features is bolded per task.}
\label{table:uar_results}
\begin{tabular}{cccrrr}
\toprule
\bm{$\mathcal{D}$} & \bm{$\mathcal{F}$} & \textbf{SR} & \textbf{CTID} &  \textbf{CLID} &  \textbf{SID} \\
\midrule
\midrule
\multirow{2}{*}{$D_1$} & \multirow{2}{*}{C22}        & 44.1        & 51.04          & 47.58          & N/A            \\
         &         & 16          & 37.72          & 34.54          & N/A            \\
\midrule
\multirow{2}{*}{$D_1$}         & \multirow{2}{*}{WLM}        & \multirow{2}{*}{16}          & 60.10          & 67.47          & N/A            \\
         &         &           & 33.74          & 36.05          & N/A            \\
\midrule
\multirow{2}{*}{$D_1$}         & \multirow{2}{*}{E2E}        & 44          & \textbf{68.32} & \textbf{74.12} & N/A            \\
         &         & 16          & 53.03          & 59.94          & N/A            \\
\midrule
\midrule
\multirow{3}{*}{$D_2$}         & \multirow{3}{*}{C22}        & 300         & 37.68          & 43.56          & \textbf{66.24}          \\
         &         & 60          & 32.50          & 35.52          & 63.38          \\
         &         & 16          & 35.65          & 35.32          & 58.14          \\
\midrule
\multirow{2}{*}{$D_2$}         & \multirow{2}{*}{WLM}        & \multirow{2}{*}{16}          & \textbf{56.77} & 46.05          & 63.80          \\
         &         &           & 32.11          & 25.42          & 57.98          \\
\midrule
\multirow{2}{*}{$D_2$}         & \multirow{2}{*}{E2E}        & 60          & 42.03          & \textbf{49.78} & 62.36          \\
         &         & 16          & 37.65          & 36.21          & 60.15          \\
\midrule
\midrule
\multirow{3}{*}{$D_3$}         & \multirow{3}{*}{C22}        & 125         & 64.32          & 43.19          & 62.80          \\
         &         & 60          & 65.67          & 45.50          & 61.22          \\
         &         & 16          & 52.59          & 39.43          & 57.32          \\
\midrule
\multirow{2}{*}{$D_3$}         & \multirow{2}{*}{WLM}        & \multirow{2}{*}{16}          & \textbf{80.38} & \textbf{55.58} & \textbf{74.26} \\
         &         &           & 64.62          & 41.33          & 59.14          \\
\midrule
\multirow{2}{*}{$D_3$}         & \multirow{2}{*}{E2E}        & 60          & 65.31          & 47.92          & 60.73          \\
         &         & 16          & 66.24          & 31.31          & 56.59          \\
\bottomrule
\end{tabular}
\end{table}

Ignoring the sampling frequency aspect, it can be observed that E2E yields the best performances for $D_1$'s CTID and CLID tasks. For $D_2$, WLM yields best performance for CTID, E2E for CLID, and C22 for SID. On both $D_1$ and $D_2$, we can observe that WLM yields competitive systems, however in the case of $D_3$, WLM's third layer representations consistently yield the best performance across all the tasks (see \Cref{fig:combined_layers}), and outperform C22 and E2E. Although WLM yields competitive performances on $D_1$ and $D_2$, it is difficult to systematically compare to C22 or E2E as different layers yield best performance for different tasks.

Furthermore, it can be observed that the 16 kHz SR performance is generally inferior across different datasets and tasks for C22 and E2E. This finding is in line with the understandings in the literature gained by analysis of different call types which showed that most marmoset call types extend into frequencies above 8 kHz \cite{Agamaite2015}. This implies that, with an 8 kHz bandwidth, certain vital information for specific call types might be lost, rendering it increasingly challenging, if not impossible, for the classifier to accurately categorize certain calls. Indeed, it can be observed that C22 systems yield superior performance with the native SR compared to 16 kHz for all datasets. This emphasizes that higher frequencies are likely to contain valuable information. A comparison between C22, WLM and E2E at 16 kHz sampling frequency demonstrates the potential of SSL based feature representations learned on human speech.

It is worth noting that a recent, independent study explored representations learned from other acoustic domains such as general audio, which includes audio event classes such as environmental sounds, musical instruments, and human and animal vocalizations. They demonstrated on $D_1$ that increasing the pre-training bandwidth of a PANN model \cite{PANN}, pre-trained on the AudioSet dataset with log-mel spectrogram inputs, improved performance on both CTID and CLID tasks \cite{sarkar24_vihar}. However, the study didn't explicitly disentangle whether these improvements resulted from the increased bandwidth itself, the spectrogram-based inputs, or from the inclusion of some animal vocalizations in the pre-training dataset. This distinction still remains an important open question for future investigations.

\section{Analysis}
\label{sec:analysis}

\subsection{Layer-wise linear performance analysis}
\label{ssec:layerwise_exp}
\begin{figure}[ht]
  \centering
  \includegraphics[width=\linewidth]{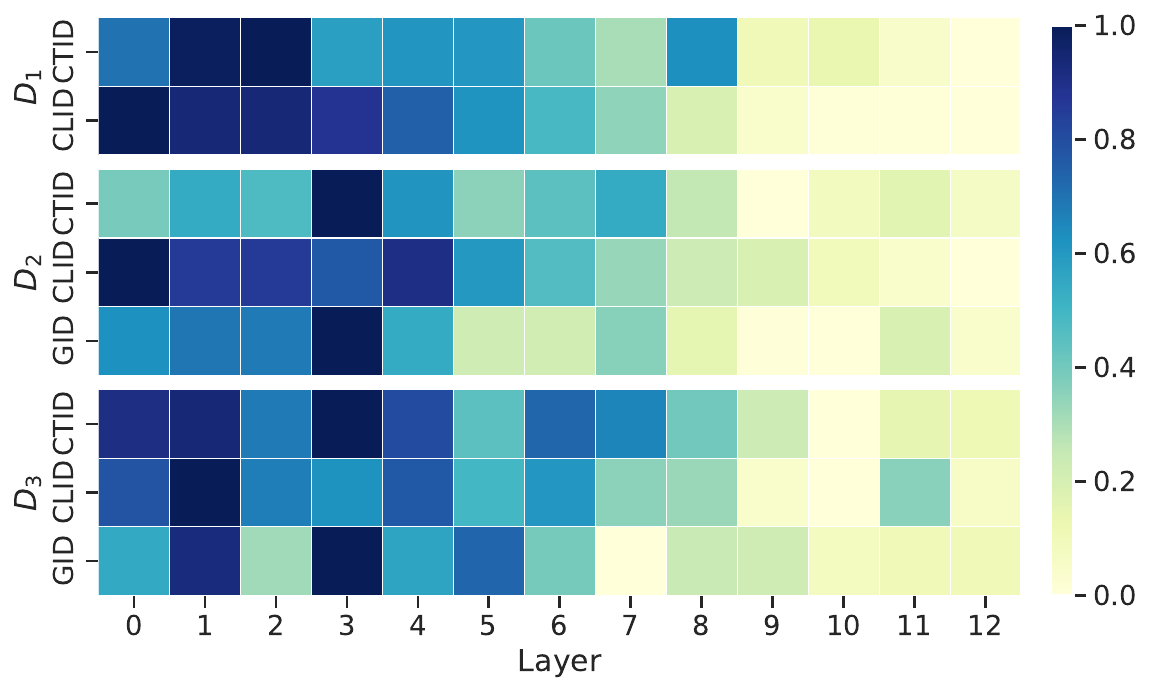}
  \caption{Layer-wise UAR scores of WLM features modeled by single layer perceptron. The scores are normalized independently per task. Darker regions indicate higher performance.}
  \label{fig:layer_matrix}
\end{figure}

In \Cref{fig:combined_layers}, it can be observed that lower layer representations tend to yield better systems. To further ascertain that, we carried out layer-wise classification performance of the same tasks using a simple linear classifier (single layer perceptron). \Cref{fig:layer_matrix} shows the results independently normalized per-task to a [0, 1] range. It can be observed that the lower layers are much more salient representations for all three tasks across all datasets when compared to higher layers. A possible explanation is that, because WavLM's CNN encoder operates directly on the raw waveform, the early layers capture fundamental \textit{acoustic} features and can leverage spectro-temporal variations relevant to tasks such as speaker identification and verification \cite{wavlm_paper}. Thus, these lower layers inherently generalize better to other acoustic domains, such as marmoset vocalizations. In contrast, the later layers -- shown to perform well on \textit{linguistic} tasks, such as speech or phoneme recognition -- appear more specialized for human speech and consequently much less transferable to bioacoustics, resulting in lower performance. We can also observe that there is no consistent optimal layer for each task type across the datasets.

\subsection{Frequency response of learnt convolution filters}
\label{ssec:learnt_filters}
We analyzed the frequency response of the first learnt convolution layer filters of E2E systems by estimating the cumulative frequency response $F_{cum}$ as~\cite{Palaz_SPEECHCOMMUNICATION_2019}:
\begin{equation} \label{cum_freq}
    F_{cum} = \sum_{k=1}^{n_f} \frac{F_k}{ \Vert F_k \Vert}_2,
\end{equation}
where $n_f$ denotes the 128 filters in the first convolution layer (see Appendix) and $F_k$ denotes discrete Fourier transform of filter $k$ over 2048 DFT points.

\begin{figure}[ht]
  \centering
  \includegraphics[width=\linewidth]{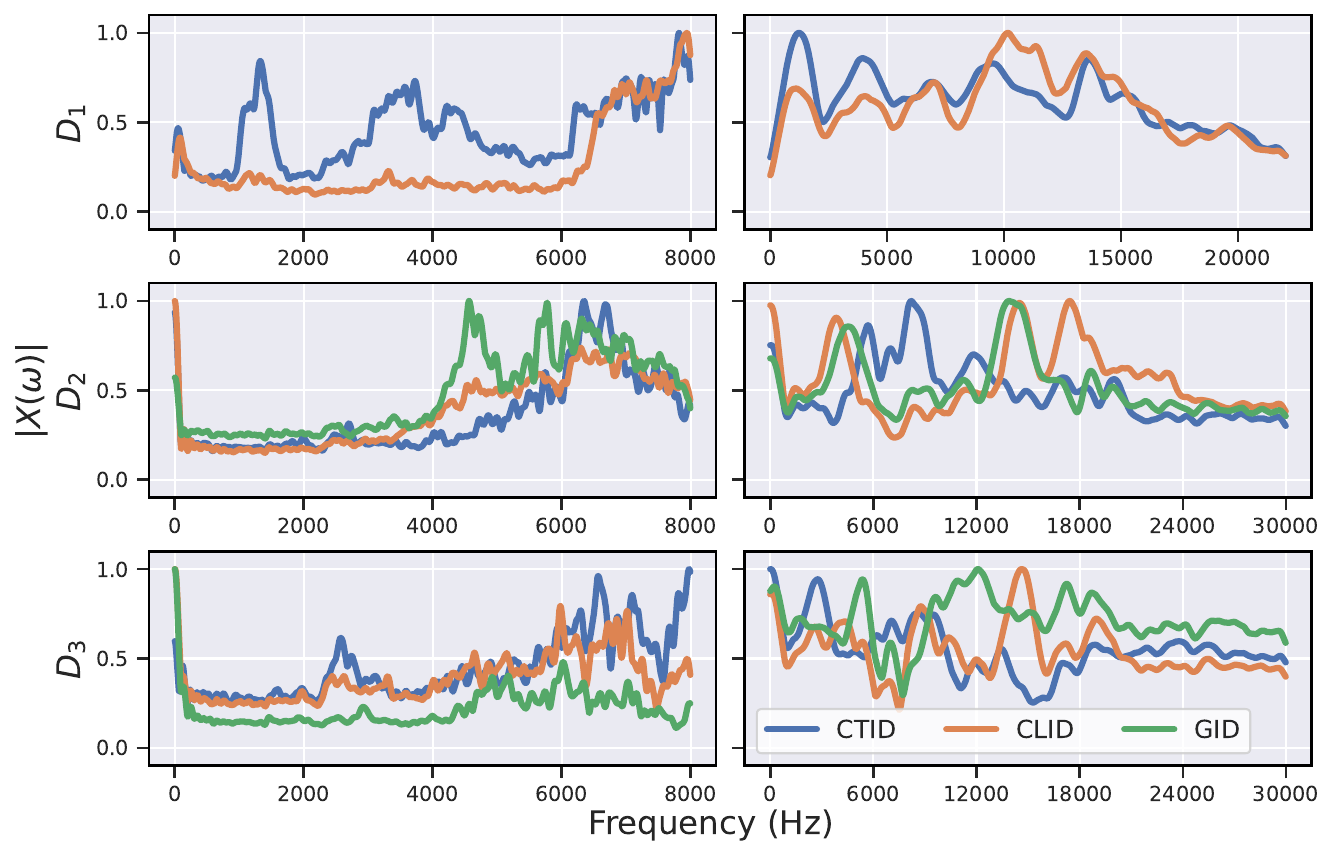}
  \caption{Cumulative frequency response per task on all datasets. Sampling rate: 16 kHz (left), and 44.1 or 60 kHz (right).}
  \label{fig:freq_reponse}
\end{figure}

\Cref{fig:freq_reponse} shows the cumulative frequency response for each task per dataset at an SR of 16 kHz, and 44.1 or 60 kHz. With a 8 kHz bandwidth (left half), it can be observed that the emphasis is on frequencies 4-5 kHz and above irrespective of the task. As the bandwidth of the signal is increased (right half), it can be observed that emphasis is also given to higher frequency regions such as around 10 kHz or above. These observations further corroborate previous findings that most marmoset calls occupy frequency ranges beyond 8 kHz~\cite{Agamaite2015}, and also explain the improved performance obtained with higher bandwidth signals. In addition, we observe that for different tasks the learned filters give emphasis to different frequency regions. A detailed analysis  of the spectral information learned is part of our future work. Taken together, the analysis indicates that the E2E framework inspired from speech processing can be scaled to marmoset call analysis.

\section{Conclusions}
\label{sec:conclusion}
This paper explored different feature representations or learning methods, namely handcrafted feature Catch22, SSL feature representation WLM, and end-to-end acoustic modeling (E2E) for analyzing marmoset calls. Our investigations on three different datasets demonstrate that end-to-end acoustic modeling and SSL feature representations yield better systems than handcrafted Catch-22 features for call-type classification and caller identification, while also achieving comparable performances for sex identification at a common sampling rate. As a by-product, our studies demonstrated that (a) the utility of pre-trained SSL models on human speech can be extended to call-type and sex, besides caller discrimination and (b) end-to-end acoustic modeling methods developed for speech processing can be scaled for marmoset call analysis. Our study raises a few pertinent questions such as: (a) with limited signal bandwidth how are SSL features informative about marmoset calls? (b) what kind of task specific spectral information is learned by the E2E systems?, and (c) how to combine the different approaches for improving marmoset call analysis? Furthermore, in this work we only investigated feature representations that directly modeled the raw input waveform. However, recent bioacoustic studies on bats, birds, and rodents have leveraged spectrogram-based methods \cite{goffinet, Ruff, Coffey, 10447620}. Whether such approaches can offer distinct advantages over the waveform-based methods for marmoset vocal communication analysis remains to be determined. Our future work will investigate these questions.

\section*{Data Availability}
\noindent The Dataset $D_1$ and a corresponding PyTorch Dataloader is publicly available on Zenodo\footnote{\href{https://zenodo.org/records/10130104}{\texttt{https://zenodo.org/records/10130104}}\label{note_zen}}, with reference number \texttt{10130104}. In addition, the datasets $D_2$ and $D_3$ are both available from the corresponding authors upon request.

\section*{CRediT Authorship Contribution Statement}
The following paragraph details the CRediT (Contributor Roles Taxonomy) designations for each author:

\textbf{ES}: Conceptualization, Data Curation, Methodology, Investigation, Formal Analysis, Writing – original draft Preparation, Writing – review and editing, Resources. \textbf{KW}: Data curation, Formal Analysis, Writing – review and editing, Resources. \textbf{ABB}: Data Curation, Writing – review and editing, Resources. \textbf{JB}: Writing – review and editing, Supervision, Resources, Funding acquisition. \textbf{MMD}: Conceptualization, Methodology, Writing – original draft, Writing – review and editing, Supervision, Resources, Funding acquisition.

\section*{Declaration of Competing Interest}
The authors declare that they have no known competing financial interests or personal relationships that could have appeared to
influence the work reported in this paper.

\section*{Generative AI Disclosure Statement}
The authors acknowledge using OpenAI’s ChatGPT (models \texttt{GPT-4o} and \texttt{o1}) solely for language editing and improving readability. All results, analyses, interpretations, conclusions, and overall research findings presented in this paper remain exclusively the original work of the authors.

\section*{Acknowledgements}
This work was funded by the Swiss National Science Foundation's (SNSF) National Centre of Competence in Research (NCCR) Evolving Language project (grant 51NF40\_180888).

\appendix
\section{Data description}
\label{sec:data_description}
Dataset $D_1$ is an extended version of the dataset used in the study on marmoset call type discrimination by Zhang et al.~\cite{cas_data}. This version, entitled InfantMarmosetsVox, was used in the recent work on marmoset caller discrimination using SSL features~\cite{Sarkar_INTERSPEECH_2023}. The audio was recorded from five pairs of infant marmoset twins, each recorded individually in two separate sound-proofed recording rooms at a sampling rate of 44.1 kHz. Additionally, marmosets were recorded individually without communication with other marmosets and the intervention from experimenters. The audio recordings were manually annotated using the Praat tool by an experienced researcher. For each vocalization, the start and end time, call type, and marmoset identity have been provided. The data consists of 11 different marmoset calltypes, namely, peep (pre-phee), phee, twitter, trill, trillphee, tsik tse, egg, pheecry (cry), trllTwitter, pheetwitter, and peep. The data contains 350 files of precisely labelled 10-minute audio recordings across all ten caller classes.

$D_2$ consists of 102 labelled 10-min focal audio recordings of common marmoset calls recorded in six behavioural contexts. A pair of marmosets was either separated or in the same enclosure, with preferred food either freely available for the focal individual or not. Each of the 8 subjects was recorded on 16 separate occasions. Most of the calls were given in bouts as holistic single call units, and thus, a call-type unit was defined as a call bout with call elements which were not further apart than 0.5s, as per existing literature \cite{Agamaite2015, Snowdon2001}. We only used the segments labelled as single call elements, i.e. not split up in bouts, to avoid data overlap and duplication. The dataset consists of 7 calls, namely alarm, ek, food, phee, trill, tsk, and twitter. The audio recordings were manually annotated by using Avisoft SASLab Pro (Avisoft Bioacoustics, Feb. 2017) to narrowly label the start and end of each call-type. The data was collected under Swiss legislation and licensed by Zurich's cantonal veterinary office (license ZH 223/16 and ZH 232/19).

$D_3$  was collected from 6 target adult common marmosets, 3 male and 3 female, housed at the University of Zurich. Two additional non-target individuals were also included in the dataset, summing to 8 individuals in total. The data consists of 12 calls classes: phee, trill, food call, tsk, low tsk (tsk with a peak frequency of approximately 7-9 kHz), twitter (sequence), ek, phee sequence (multiple phees), low tsk sequence (multiple low tsks), ek sequence (multiple eks), food call sequence (multiple food calls). All procedures were done in accordance with Swiss legislation and were licensed by Zurich’s cantonal veterinary office (license ZH223/19). For each recording, two individuals (one male and one female) were placed in adjacent wire cages and recorded simultaneously in 15-minute intervals with two UltraSoundGate 116H recorders coupled with an Avisoft CM16/CMPA condenser microphone (Avisoft Bioacoustics, Germany), each set to a different gain to capture both low and high amplitude calls with a sampling rate of 125kHz. A total of 12 recordings, spread over 7 months, were made for each target individual. Caller identity was labeled in real time using Avisoft-RECORDER USGH (Avisoft Bioacoustics, Germany). The labelling of the calls’ exact start and end points was carried out through a visual examination of the spectrograms. For inclusion in subsequent analyses, calls needed be distinctly visible on the spectrogram, devoid of any interference from other calls, and readily classifiable into specific call-type categories.

\section{CNN architecture}
\label{sec:cnn_arch}

\Cref{table:CNN} presents the architecture of the E2E system. The first convolution layer kernel width $kW$ and shift $dW$ was chosen based on the sampling frequency. More precisely, based on the understanding gained from speech studies, we chose those hyper-parameters to strike a balance between the length of the convolution filter and enough pitch cycles being modeled~\cite{hannah_verif_2018}. For 44.1 and 60 kHz sampling frequency, we chose $kW = 1$ ms and $dW = 0.05$ ms, respectively. As marmoset calls have fundamental frequency around 5 kHz  and above~\cite{Agamaite2015}, 1 ms signal would be expected to contain around 10 pitch cycles or more. However, for 16 kHz sampling frequency, 1 ms would contain only 16 samples, i.e. at the most 1-2 sample(s) representing each pitch cycle. This may not hinder capturing the pitch frequency information in the marmoset call well. So, for 16 kHz we set $kW = 10$ ms and $dW = 0.5$ ms.  The training batch size 16 and learning rate of 0.001, same as the MLP classifier for C22 and WLM. The optimization configuration simply consisted of Adam and a dynamic learning rate scheduler which reduces the learning rate $\eta$ when the selected optimization criterion, in this case \textit{Val} UAR, shows no improvement after 10 epochs.

\begin{table}[!htb]
\centering
\caption{CNN model parameters. $n_f$ denotes the number of filters, $n_{hu}$ the the number of hidden units, and $\sigma$ the activation function.}
\begin{tabular}{lccccc}
\toprule
\textbf{Layer} & \bm{$kW$} & \bm{$dW$} & \bm{$n_f$}/\bm{$n_{hu}$} & \textbf{Padding} & \bm{$\sigma$} \\
\midrule
Conv 1 & \textit{kW}  & \textit{dW}     & 128   & -   & ReLU \\
Conv 2 & 10  & 5     & 256   & -   & ReLU \\
Conv 3 & 4   & 2     & 512   & 2   & ReLU \\
Conv 4 & 3   & 1     & 512   & 1   & ReLU \\
Adapt  &  -  & -     &  -    & -   & -   \\
FC 1   &  -  & -     & 512   &  -  & ReLU \\
FC 2   &  -  & -     & 256   &  -  & ReLU \\
FC 3   &  -  & -     & $n_c$ &  -  & - \\
\bottomrule
\end{tabular}
\label{table:CNN}
\end{table}

\bibliographystyle{elsarticle-num} 
\bibliography{ref}

\end{document}